\documentclass[multphys,vecphys]{svmult}
\usepackage{graphicx,epsfig,makeidx,multicol}
\usepackage[bottom]{footmisc}
\usepackage{epstopdf}
\usepackage{bm}
\usepackage{latexsym,amssymb,amsmath,mathrsfs}
\usepackage{url}
\usepackage[square,comma,sort&compress]{natbib}

\newcommand\La{\ensuremath{\Lambda}}

\newcommand\Ia{\ensuremath{{\cal I}}}

\newcommand\BC{\begin{center}}
\newcommand\EC{\end{center}}

\begin{document}

%
%
%
%
%
%
%
%
%
%
%
%

\title*{Island Cosmology}
\titlerunning{Island Cosmology} 
\author{Sourish Dutta\inst{1}\and
Tanmay Vachaspati\inst{2,}\inst{3}}
\authorrunning{Sourish Dutta and Tanmay Vachaspati} 
\institute{Department of Physics and Astronomy, Vanderbilt University, \\
6301 Stevenson Center, Nashville, TN 37235, USA, 
\texttt{sourish.dutta@vanderbilt.edu}
\and Institute for Advanced Study, Princeton, NJ 08540, USA
\and CERCA, Department of Physics, Case Western Reserve University, \\ 
10900 Euclid Avenue, Cleveland, OH 44106-7079, USA
\texttt{tanmay@monopole.phys.cwru.edu}}

\maketitle

\begin{abstract}
\noindent
If the observed dark energy is a cosmological constant, the canonical
state of the universe is de Sitter spacetime. In such a spacetime,
quantum fluctuations that violate the null energy condition will
create islands of matter. If the fluctuation is sufficiently large,
the island may resemble our observable universe. Phenomenological
approaches to calculating density fluctuations yield a scale invariant
spectrum with suitable amplitude. With time, the island of matter that
is our observable universe, dilutes and re-enters the cosmological
constant sea, but other islands will emerge in the future, leading
to an eternal universe.
 
\end{abstract}

\section{Introduction}
\label{introduction}

Current observations indicate a bleak future for the universe.
The expansion of the universe will accelerate and, if the dark
energy is a cosmological constant, all the matter will dilute
away, eventually leaving behind empty de Sitter space, the
canonical state of the universe.

The classical picture of an empty de Sitter spacetime is
modified when quantum field theory effects are taken into
account since inevitable quantum fluctuations can speed up 
or slow down the rate of expansion (
There has also been work to investigate whether
quantum field theory actually causes instabilities in de Sitter
space \cite{Tsamis:1996qm, Polyakov:2007mm}).
Quantum field theoretic fluctuations that violate the null 
energy condition (NEC) on super-horizon scales can lead to 
cosmological super-acceleration, producing a super-horizon patch 
of matter that then evolves as a Friedmann-Robertson-Walker 
universe embedded within the de Sitter spacetime. 

Island cosmology is based on the idea that our observable 
universe may have originated from an NEC violating quantum 
fluctuation in the cosmological constant sea. There are a
number of attractive features of such an hypothesis. For
example, spacetime is eternal, there need not be any
singularities, there are observable islands to the future
as there are in the past, and the model does not require
new physics or unobserved ingredients. The real test of
the model is in how successfully it confronts observations
and the crucial signatures to study are the spectrum of 
predicted density and gravitational wave fluctuations.

We now summarize the main features of island cosmology,
and discuss the points in detail in the following sections.

Let us begin with a broad overview of Island Cosmology, 
describing each of its different stages.
\begin{figure}
\scalebox{0.40}{\includegraphics{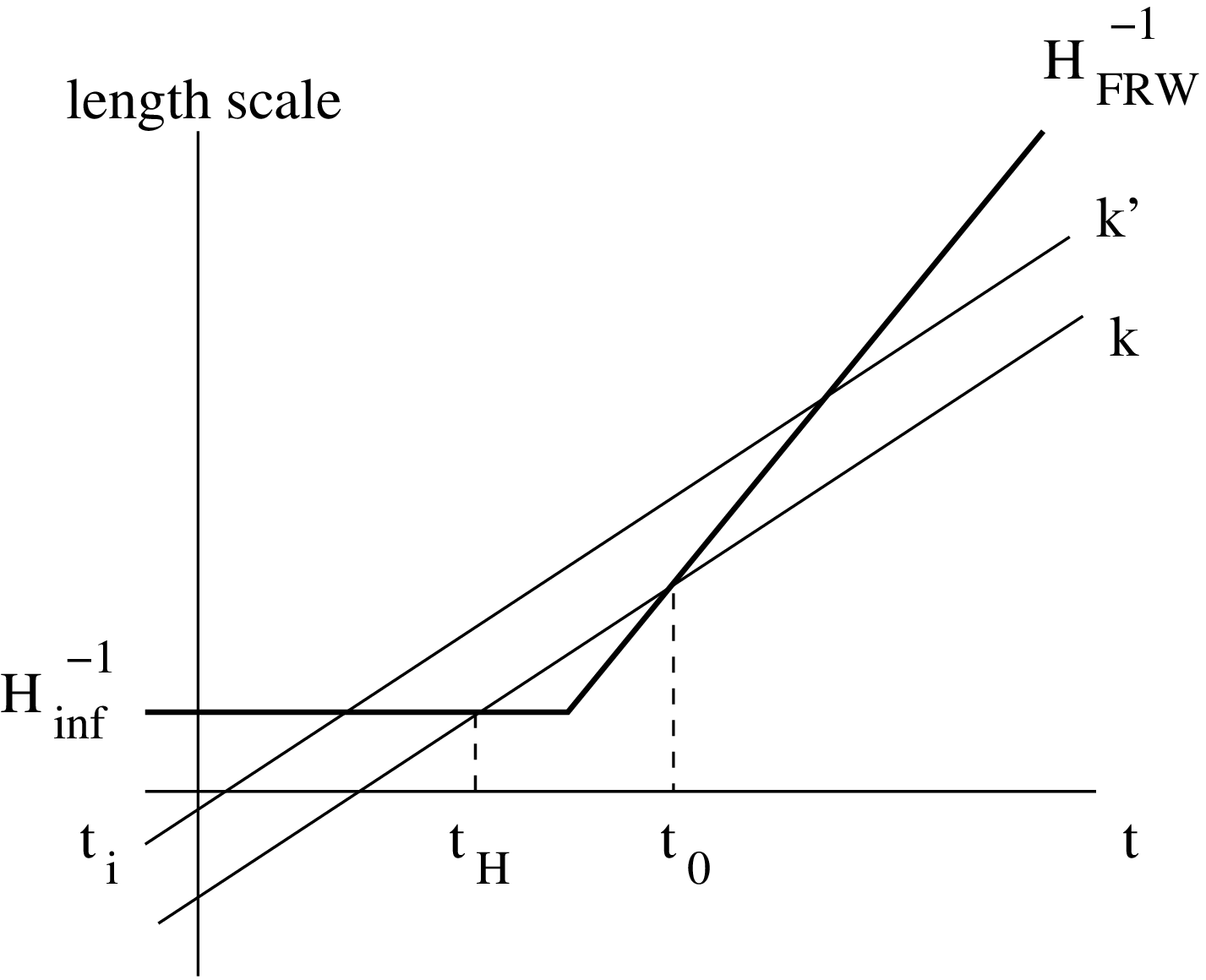}}
\scalebox{0.40}{\includegraphics{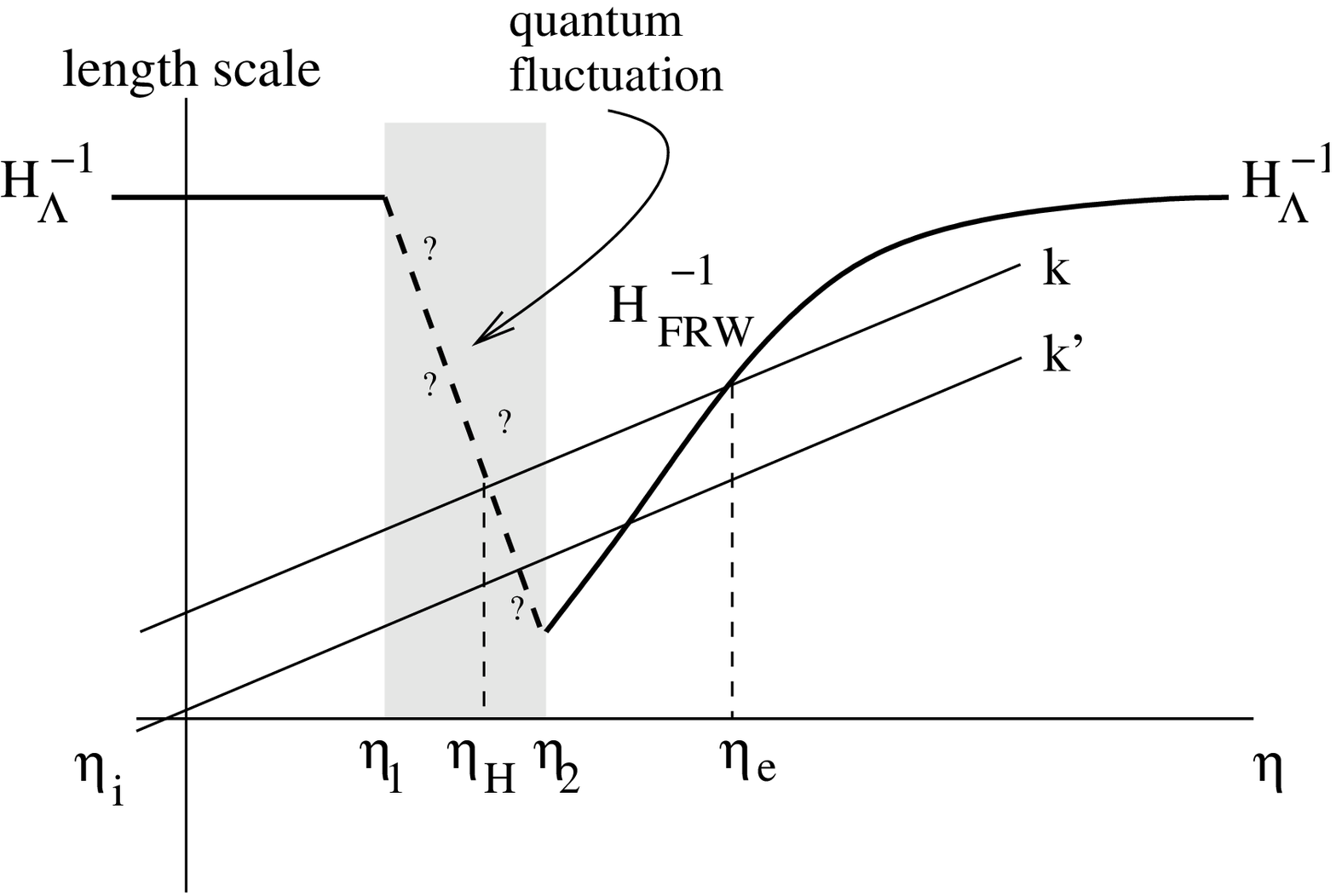}}
\caption{\label{figure} Sketch of the behavior of the Hubble length
scale in inflationary cosmology (left) with respect to cosmic time, 
$t$, and in island cosmology (right) with respect to conformal time, 
$\eta$, (which is defined in terms of the scale factor $a(t)$ as $d\eta=a(t)dt$ ). 
$H_{\text{inf}}^{-1}$, $H_{\text{FRW}}^{-1}$ and $H_{\Lambda}^{-1}$ denote 
the Hubble lengths during an inflationary, Friedman-Robertson-Walker and 
cosmological constant dominated evolution respectively. The evolution of fluctuation modes 
is also shown in the two models. In inflationary models, early 
exponential growth is driven by an inflaton field, while in island 
cosmology it is driven by the presently observed dark energy, assumed 
to be a cosmological constant. As the cosmological constant is very 
small, the Hubble length scale is very large -- of order the present 
horizon size. The exponential expansion in island cosmology ends
in some horizon volume not due to the decay of the vacuum energy 
as in inflationary scenarios but due to a quantum fluctuation in 
the time interval $(\eta_1, \eta_2)$ that violates the Null Energy 
Condition (NEC). The NEC violating quantum fluctuation  causes the 
Hubble length scale to decrease.  After the fluctuation is over, the 
universe enters radiation dominated FRW expansion, and the Hubble 
length scale grows with time. The physical wavelength of a quantum 
fluctuation mode starts out less than $H^{-1}_\Lambda$ at some early 
time $\eta_i$. The mode exits the cosmological horizon during the NEC 
violating fluctuation ($\eta_H$) and then re-enters the horizon at 
some later epoch $\eta_e$ during the FRW epoch.
}
\end{figure}

\begin{enumerate}
\item {\textbf{The $\La$-sea:}} Current observations indicate that 
the Universe is accelerating, with the acceleration driven by some 
form of Dark Energy \cite{riess-1998-116}. The simplest explanation 
for a dark energy candidate, which is consistent with observations, 
is the cosmological constant, $\Lambda$. We take such an expanding 
universe, inflating with the observed value of the cosmological constant, 
to be the starting point of our model. In other words, Island Cosmology 
begins with an expanding de Sitter spacetime with a horizon size 
comparable to our present horizon $H^{-1}_0$. We call this spacetime 
the $\La$-sea.

\item {\textbf{The Upheaval:}} A quantum fluctuation of some field 
({\it e.g.} scalar field, photon) in a horizon-size volume (which 
we call an $\cal I$-region) in the $\La$-sea drives the Hubble constant 
to a large value within this volume. We call tis fluctuation the "upheaval". 
As a result of the upheaval the Hubble scale in the $\Ia$-region 
decreases, even though the universe continues to expand. A simple 
application of the Friedmann equation shows that this can only occur 
when the null energy condition (NEC) is violated, that is
\begin{equation}
N_{\mu}N_{\nu}T^{\mu\nu}\geq0
\end{equation}
where $N$ stands for a null vector, and $T^{\mu\nu}$ represents the 
energy momentum tensor. As we discuss later, it has been demonstrated
Refs.~\cite{GutVacWin??,Winitzki:2001fc,Vachaspati:2003de}, that 
quantum field theoretic fluctuations allow for this possibility.

\item {\textbf{The Island Universe:}} After the upheaval, the Hubble 
constant within the $\cal I$-region is large, and the $\cal I$-region 
gets filled with classical radiation, as the NEC-violating field decays 
into relativistic particles. Thereafter, the $\cal I$-region evolves 
as a radiation-dominated FRW Universe, and eventually forms our observed 
Universe. We call a Universe created in this manner an 
\textbf{Island Universe}.

\item {\textbf{The $\La$-sea, again:}} With further evolution, the 
Island Universe dilutes, the $\cal I$-region is again dominated by 
the cosmological constant and we are once again left with only the 
$\La$-sea. Clearly, Island Cosmology is cyclic and the above process 
can be infinitely repeated.
\end{enumerate}

Island Cosmology does include elements of earlier work such as eternal 
inflation models, steady state models and ekpyrotic models. While 
eternal inflation models \cite{Vilenkin:1983xq,Linde:1986fd}, especially 
Garriga and Vilenkin's ``recycling universe'' \cite{Garriga:1997ef}, 
(see also the discussion in \cite{Carroll:2004pn}) use NEC violating 
quantum fluctuations in the inflaton field to drive the Hubble length 
scale to smaller values, in Island cosmology, these quantum fluctuations 
can occur in {\em any} quantum field and have to be large. In both 
inflationary cosmology and our case, the quantum fluctuation needs 
to violate the NEC. Furthermore, in both cases the back-reaction of 
the fluctuation is assumed to lead to a faster rate of cosmological 
expansion. In Steady State Cosmology \cite{BonGol48,Hoy48}, matter is 
sporadically produced by ``minibangs'' in a hypothetical C-field. 
The explosive events in Island Cosmology, on the other hand, are 
quantum field theoretic in origin and seed the matter content of 
an entire Universe. The decreasing Hubble scale is also a feature 
of the ekpyrotic cosmological model \cite{Khoury:2001wf}. However, 
in that model, the motivation for the decrease lies in extra-dimensional 
brane-world physics and results in a period of contraction of our three 
dimensional universe. Island Cosmology does not involve any brane-world 
physics, and has no contracting phase, as the Universe continues to 
expand even while the Hubble scale drops.

\section{NEC violations in de Sitter space}
\label{necindS}

In de Sitter spacetime, as well as any other spacetime, there are
fluctuations of the energy-momentum tensor, $T_{\mu \nu}$, of quantum
fields. This is simply a consequence of the fact that the vacuum, 
$|0\rangle$, is an eigenstate of the Hamiltonian but not of the 
Hamiltonian density or the energy-momentum density operator, 
${\hat T}_{\mu\nu}$. In short-hand notation:
\begin{eqnarray}
{\hat T}_{\mu\nu} |0\rangle &=& \sum 
   [(\dots )a_l a^\dag_k + (\dots )a^\dag_l a^\dag_k] ~ |0\rangle
\nonumber \\
 &=& \sum [(\dots ) |0\rangle + (\dots ) |2;k,l \rangle]
\label{Tonvac}
\end{eqnarray}
where, the ellipses within parenthesis denote various combinations of
mode functions and their derivatives; $a^\dag_k$, $a_l$ are creation
and annihilation operators and $|2;k,l\rangle$ is a two particle state.
The final expression is not proportional to $|0\rangle$, implying that 
the vacuum is not an eigenstate of ${\hat T}_{\mu\nu}$ and there will 
be fluctuations of the energy-momentum tensor in de Sitter space.

In de Sitter spacetime, the quantum vacuum state needs to be chosen.
If we require that on short distance scales the quantum fields should 
behave as they do in Minkowski spacetime, then the vacuum state is
known as the Bunch-Davies vacuum \cite{Bunch:1978yq}. 
It has been shown
\cite{GutVacWin??,Winitzki:2001fc,Vachaspati:2003de}
that quantum field theory of a light scalar field in the Bunch-Davies
vacuum in de Sitter space leads to violations of
the NEC. We now briefly summarize the general arguments behind this 
conclusion.

The first step is to construct a
``smeared NEC operator''
\begin{equation}
\hat{O}^{\rm ren}_{W}\equiv 
 \int d^{4}x\sqrt{-g}\, W\left( x; R,T\right) 
N^\mu N^\nu {\hat T}^{\rm ren}_{\mu\nu}
\label{OrenW}
\end{equation}
where $ W\left( x; R,T\right)$ is a smearing function
on a length scale $R$ and time scale $T$. The vector
$N^\mu$ is chosen to be null, and the superscript
$ren$ denotes that the operator has been suitably renormalized.
As shown in \cite{GutVacWin??} the smeared operator will not be 
proportional to the vacuum state either, and will fluctuate. The 
root-mean-squared (rms)
scale of the fluctuations, $\hat{O}_{\rm rms}^2$ can be estimated on 
dimensional grounds:
\begin{equation}
\hat{O}_{\rm rms}^2 \equiv 
\langle 0 | ({\hat O}^{\rm ren}_W)^2 | 0 \rangle \sim H_\Lambda^8
\label{Oren2expec}
\end{equation}
in the special case when $R = T = H_\Lambda^{-1}$.
Since, in de Sitter space,
$\langle 0|{\hat T}_{\mu \nu} |0\rangle \propto g_{\mu\nu}$,
we also have:
\begin{equation}
\langle 0 | {\hat O}^{\rm ren}_W |0\rangle =0
\label{Orenexpec}
\end{equation}
Therefore the fluctuations of ${\hat O}^{\rm ren}_W$ are both positive 
and negative. Assuming a symmetric distribution, we come to the 
conclusion that quantum fluctuations of a scalar field violate the 
NEC with 50\% probability. Exactly the same arguments can be applied 
to quantum fluctuations of a massless gauge field such as the photon.

Note that the above calculation does not give us the probability 
distribution of the violation amplitude, for which we would
have to calculate the actual probability distribution for the operator
${\hat O}^{\rm ren}_W$. However, by continuity we can expect that large
amplitude NEC violations will also occur with some diminished but
non-zero probability.

\section{Extent and duration of NEC violation}
\label{extentandduration}

\begin{figure}
\BC\scalebox{0.80}{\includegraphics{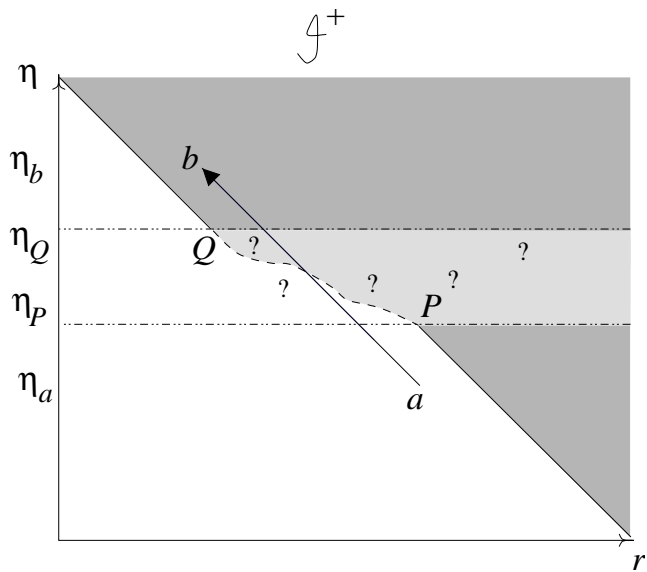}}\EC
\caption{\label{nec_violation}
We show a Penrose diagram ($\eta$ vs $r$, where $\eta$ is 
conformal time and $r$ is 
the radial coordinate of a null ray) for a classical de Sitter 
spacetime for conformal time
$\eta < \eta_P$, that transitions to a faster expanding
classical de Sitter spacetime for $\eta > \eta_Q$.
Future null infinity, denoted by $\mathscr{I}^+$, occurs along
the horizontal upper edge of the diagram.
The inverse Hubble size is shown by the white region.
A bundle of ingoing null rays originating at point $a$ is
convergent initially but becomes divergent in the superhorizon
region at point $b$. This can only occur if the null energy condition violating
is violated in the region $\eta \in (\eta_P, \eta_Q)$.
In the quantum domain, a classical picture of spacetime may
not be valid and this is made explicit by the question marks.
}
\end{figure}

What is the spatial and temporal extent of these NEC-violating 
fluctuations? Clearly, such fluctuations can occur on all spatial 
and temporal scales, but based on causality and predictability, 
we will now argue that only fluctuations of a large (horizon-sized) 
spatial extent and small temporal duration are relevant to creating 
islands of matter.

Consider the spacetime diagram of Fig.~\ref{nec_violation}.
In this diagram we show an initial de Sitter space that
later has a patch in which the space is again de Sitter
though with a larger expansion rate. Consequently, the initial
Hubble length scale $H_i^{-1}$ is larger than the final
Hubble length scale $H_f^{-1}$. It follows, therefore, that there 
are ingoing null rays that are within the horizon initially but 
eventually propagate outside the horizon. An example of such
a null ray is the line from $a$ to $b$. At point $a$ a bundle
of such rays will be converging whereas at point $b$ the
bundle will be diverging. It can be demonstrated from the Raychaudhuri 
equation (subject to a few mild conditions, such as general relativity 
being valid and spacetime topology being trivial) that the transition 
from convergence to divergence of a bundle of null rays can only occur 
if there is a NEC violation somewhere along the null ray 
(see \cite{Vachaspati:1998dy}).

\begin{figure}
\BC\scalebox{0.80}{\includegraphics{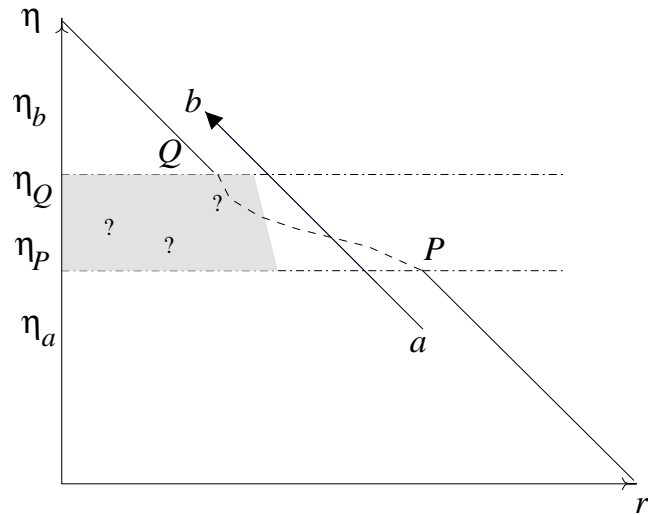}}\EC
\caption{\label{nec_patch}
A spacetime diagram similar to that in Fig.~\ref{nec_violation}
but one in which the Null Energy Condition (NEC)
violation occurs over a sub-horizon region
(shaded region in the diagram).
Now the null ray bundle from $a$ to $b$ goes from being converging
(within the horizon) to diverging (outside the horizon). However, it
does not encounter any null energy condition violation along its path, and this is not
possible as can be seen from the Raychaudhuri equation.
Since the ingoing null rays are convergent as far out as the point
$P$, the size of the quantum domain has to extend out to at
least the inverse Hubble size of the initial de Sitter space.
Therefore the NEC violating patch has to extend beyond the initial
horizon.
}
\end{figure}

Now if the NEC violation only occurred on a scale smaller than 
$H_i^{-1}$, one could imagine a null ray that would never enter 
the NEC violating region and yet go from being converging to 
diverging (see Fig.~\ref{nec_patch}). This would clearly be inconsistent
with the Raychaudhuri equation.

Furthermore, after the energy condition violations are over, the 
faster expanding region would have to either instantly revert to 
the ambient expansion rate, or some spacetime feature, such as a 
singularity, would have to occur to prevent a null ray from entering 
the faster-expanding region from the slower-expanding region. Additional 
boundary conditions would have to be imposed on the singularity to 
restore predictability. An example of such a process can be found in 
Ref.~\cite{Borde:1998wa} in connection with topological inflation 
\cite{Vilenkin:1994pv,Linde:1994hy}.

Another way of understanding the loss of predictability is the 
following. Whenever a faster expanding universe is created, it must 
be connected by a wormhole to the ambient slower expanding region. 
The wormhole can be kept open if the energy conditions are 
violated \cite{Morris:1988cz}. But, if the wormhole neck is small, 
as soon as the energy condition violations are over, it must collapse 
and pinch off into a singularity. Signals from the singularity can 
propagate into the faster expanding universe destroying predictability. 
However, if the neck of the wormhole is larger than the horizon size 
of the ambient universe, the ambient expansion can hold up the wormhole 
and the neck does not collapse even after the NEC violation is over.

Our argument that NEC violations on scales larger than
the horizon are needed to produce a faster expanding universe is
consistent with earlier work \cite{Farhi:1986ty} showing that it is
not possible to produce a universe in a laboratory without an initial
singularity (also see \cite{Linde:1991sk}). Subsequent discussion of
this problem in the quantum
context \cite{Fischler:1989se,Farhi:1989yr,Fischler:1990pk}, however,
showed that a universe may tunnel from nothing
without an initial singularity, just as in quantum cosmology
\cite{Vilenkin:1982de,Hartle:1983ai}. Such a tunneling event, however, 
is irrelevant to Island Cosmology, as the newly
created universe is causally disconnected from the ambient $\Lambda$-sea.
Without an inflaton, the process would therefore produce only a second
$\Lambda$-sea.

Based on the above arguments, and on the results of the earlier 
investigations cited, we conclude that to get a faster expanding region 
that lasts beyond the duration of the quantum fluctuation and remains
predictable, the spatial extent of the NEC violating fluctuation
must be larger than
$H_i^{-1}$:
\begin{equation}
R > H_i^{-1}
\end{equation}
where $R$ is the spatial extent of the fluctuation and shows
up as the spatial smearing scale in the calculation of ${\hat O}_{\rm rms}$.

We also argue that the temporal scale of the fluctuation has to be small. 
This is because, an explicit evaluation \cite{GutVacWin??} shows that 
${\hat O}_{\rm rms}$ is proportional to inverse powers of the temporal 
smearing scale and diverges as the
smearing time scale $T \rightarrow 0$. Hence the briefer the fluctuation,
the stronger it can be, as we might also expect from an application of
the Heisenberg time-energy uncertainty relation. Therefore we
take the time scale of the NEC violation to be vanishingly
small:
\begin{equation}
T \rightarrow 0
\end{equation}

\section{Likelihood -- the role of the observer}
\label{anthropics}

Any cosmology that relies on fluctuations or initial conditions
must explain why the particular fluctuation or initial condition
was chosen. This issue is common to {\it all} presently known
cosmological models. For example, inflationary cosmology chooses 
the inflaton in a particular location on the potential as an initial 
condition (``top of the hill'') and a starting universe that consists 
of several horizons with homogeneous conditions. In island cosmology
too, we need to discuss the likelihood of having NEC violating
fluctuations that are large enough to produce habitable islands.

In Sec.~\ref{extentandduration} we have pointed out that the 
NEC-violating fluctuations need to have two requirements to be 
cosmologically relevant - they need to have a superhorizon spatial 
extent and must be of vanishingly small duration. There is one more 
requirement that is absolutely crucial - the fluctuations must have 
the correct amplitude in order to have sufficient energy density to 
lead to our observed Universe. Clearly, only if the temperature produced 
is high enough and the end point of the NEC violating
fluctuation is a thermal state with all the different forms of matter
in thermal equilibrium, further evolution of the island will simply follow
the standard big bang cosmology.

Admittedly the three requirements of large spatial extent, small 
temporal extent and large amplitude make these fluctuations rare.  
However, since spacetime is eternal in this model, we can wait 
indefinitely for such a fluctuation to occur.
The probability of fluctuations in the $\Lambda$-sea that can lead to
an inflating cosmology versus those that produce an FRW universe have
been considered by several researchers \cite{Dyson:2002pf,Albrecht:2004ke}.
In particular, Dyson et al. \cite{Dyson:2002pf} estimate probabilities 
based on a ``causal patch'' picture, which assumes that the physics 
beyond the de Sitter horizon is irrelevant to the physics within, and 
that the latter should be regarded as the complete physics of the Universe. 
Based on this picture, the authors of \cite{Dyson:2002pf} conclude that 
it is much more probable to directly create a universe like ours than 
to arrive at our present state via inflation. Albrecht and Sorbo 
\cite{Albrecht:2004ke} have argued that the conclusion rests crucially 
on the causal patch picture, and provide a different calculation leading 
to the conclusion that inflationary cosmology is favored. Both the above 
calculations assume the existence of fields that are suitable for 
inflation. However, Island Cosmology does not rely on the hypothesis 
of the inflaton, and so the comparison of the likelihood of inflation 
versus no inflation is moot.

The monopole overabundance problem can be resolved in Island Cosmology 
in a manner similar to that proposed in Ref.~\cite{Dvali:1995cj}, by 
assuming that the temperature required for magnetic monopoles production 
is higher than that required for matter-genesis. If the temperature at the
beginning of the FRW phase is below that needed for monopole
formation but above the matter-genesis temperature then there will
be no cosmological magnetic monopole problem.

Another important question is where we are located on the island. Are 
we close to the edge of the island (``beach'')? In that case we would 
observe anisotropies in the CMB since in some directions we would see 
the $\Lambda$-sea while in others we would see inland.
However, the island is very large (by a factor $a_0/a_f$, the ratio of the 
scale factors today and at the end of the NEC violation) compared to
our present horizon, $H_\Lambda^{-1}$. If we assume a uniform probability
for our location on the island, our distance from the $\Lambda$-sea will
be an $O(1)$ fraction of $H_\Lambda^{-1} a_0/a_f$. Since $a_0 /a_f$ is
of order $T_{mg}/T_0$ -- the ratio of the matter-genesis temperature
to the present temperature --  we are most likely to be sufficiently
inland so as not to observe any anisotropy in the CMB.

\section{The NEC violating field}
\label{whichfield}

Whereas inflationary models crucially rely on the existence of a
suitable scalar field (inflaton), we have so far not specified
the quantum field that causes the NEC violating fluctuation.
We now turn to this issue.

During the NEC-violating fluctuation, the energy density of the 
universe satisfies $\rho +p < 0$, or in other words, behaves like 
a phantom field (a field with an equation of state parameter $w<-1$).
In addition, in order to be able to compute the spectrum of 
perturbations arising from this model, we make the  assumption that the 
backreaction is given by the Friedmann equation, which requires that 
$\rho > 0$. Hence we need a quantum field whose energy density 
has to be positive, but whose pressure should be sufficiently negative 
so that the NEC is violated.

First consider a scalar field, $\phi$, with potential $V(\phi )$. The
energy density and pressure are:
\begin{eqnarray}
{\hat \rho} &=& \frac{1}{2} {\dot \phi}^2 + \frac{1}{2} ({\bm \nabla} \phi )^2 
        + V(\phi ) \nonumber \\ 
{\hat p} &=& \frac{1}{2} {\dot \phi}^2 - \frac{1}{6} ({\bm \nabla} \phi )^2 
        - V(\phi ) 
\end{eqnarray}
where the hats on $\rho$ and $p$ emphasize that these are quantum
operators Therefore:
\begin{equation}
{\hat \rho} + {\hat p} = {\dot \phi}^2 + \frac{({\bm \nabla} \phi )^2}{3} 
\end{equation}

The operators ${\hat \rho}$ and ${\hat \rho}+{\hat p}$ are not
proportional to each other and fluctuations in one do not have
to be correlated with fluctuations of the other. The energy
density in a region can be positive while the NEC is violated.
Therefore a scalar field, even if $V(\phi ) =0$, can provide suitable
NEC violating fluctuations.

Particle physics in the very early stages of the model is described 
by low energy particle physics that we know so well. At present we 
do not have any experimental evidence for a scalar field. The only 
light field that we know of today is the electromagnetic field. Could 
the electromagnetic field give rise to a suitable NEC violating 
fluctuation?

For the electromagnetic field, the energy density
and pressure can be written terms of the electric and
magnetic fields ${\bf E}$  and ${\bf B}$ respectively as:
\begin{eqnarray}
{\hat \rho} = \frac{1}{2} ( {\bf E}^2 + {\bf B}^2 ) \nonumber \\ 
{\hat p} = \frac{1}{6} ( {\bf E}^2 + {\bf B}^2 ) = \frac{1}{3} {\hat \rho}
\end{eqnarray}
So now ${\hat \rho}$ and ${\hat p}$ are not independent operators and
\begin{equation}
{\hat \rho} + {\hat p} = \frac{4}{3} {\hat \rho}
\end{equation}
From this relationship between the operators, it is clear that
the only electromagnetic fluctuation that can violate the NEC also
has negative energy density. This means that even though the
electromagnetic field can violate the NEC, it does not satisfy
the positive energy density condition when it does violate the
NEC, and this makes it hard to find the backreaction of the
fluctuation on the spacetime metric. It may be possible that the 
electromagnetic field will still be found to be suitable once we 
know better how to handle the backreaction problem. Then perhaps 
we will not need to rely on the classical cosmological equations
that require positive energy density.

There is a possible loophole in our discussion of the electromagnetic
field. The equation of state ${\hat p} = {\hat \rho}/3$ follows from the
conformal invariance of the electromagnetic field ${\hat T}^\mu_\mu =0$.
However, we know that quantum effects in curved spacetime give rise
to a conformal anomaly and the trace $\langle {\hat T}^\mu_\mu \rangle$
is not precisely zero. So we can expect that the equation of state
${\hat p} = {\hat \rho}/3$ is also anomalous. Whether this anomaly
can allow for NEC violations with positive energy density is not clear
to us.

Note that it is not necessary for the NEC violation to originate from
a fluctuation of a massless or light field. The arguments of
Sec.~\ref{necindS} are very general and apply to massive fields
as well.  Though, for a massive field, ${\hat O}_{\rm rms}$ will be
further suppressed by exponential factors whose exponent depends
on powers of $H_\Lambda /m$.

\section{The perturbation spectrum}

One of the most crucial observational tests for any cosmology is the
spectral index of perturbations that it generates.

Unlike most models of inflation, computing the perturbation spectrum 
in Island Universes is extremely difficult, for two reasons. Firstly, 
one requires an adequate characterization of the back-reaction of the 
NEC violating quantum fluctuations on the metric. The second complication 
arises from the fact that both the background field $\phi$ and its 
perturbation $\delta\phi$ are quantum operators, unlike the case of 
inflation, where the background field is the solution to some
classical equation of motion.

If the backreaction can still be described by the Friedmann equation, 
it has been shown that perturbation spectra of fields other than, and 
not interacting with, the NEC violating field, have a nearly scale-invariant 
spectrum \cite{Dutta:2005gt}. Assuming that the NEC-violating field
itself behaves as a classical phantom field for the duration of the NEC
violation, it  has been shown \cite{Dutta:2005if} that both the scalar and
tensor spectra are scale-invariant (also see \cite{Piao:2005na}). In
the latter case, interestingly, the scalar spectrum turns out to have
an amplitude of $H_{f}/m_{Pl}$, (where $H_f$ is the value of the Hubble 
parameter at the end of the fluctuation, and $m_{Pl}$ is the Planck mass)
implying that if the NEC violation ends at the GUT scale, then the 
amplitude of fluctuations is sufficient to seed structure. A full 
quantum-mechanical treatment of this problem is necessary.

\section{A critical review of Island Cosmology}
\label{assumptions}

Finally, we put Island Cosmology in perspective by taking a 
critical look at the assumptions upon which it is based, and 
comparing these to the assumptions underpinning other cosmological 
models.

Our first assumption is that the dark energy is a cosmological
constant. This is consistent with observations and moreover is
the simplest explanation of the Hubble acceleration. We assume
that the cosmological constant provides us with a background
de Sitter spacetime that is eternal (For a discussion
of the timescale on which the spacetime can remain de Sitter,
see Ref.~\cite{Goheer:2002vf}). As de Sitter spacetime also
has a contracting phase, the singularity theorems of
Ref.~\cite{Borde:2001nh} are evaded.

The second assumption is the existence of a field in the model
responsible for the NEC violation. A scalar field seems to be the 
likeliest candidate, though  an electromagnetic field would have 
been more satisfactory. However, we have shown (up to the loophole 
of the conformal anomaly) that the conformal invariance of the 
electromagnetic field prevents NEC violations with positive energy 
density. It is possible that with a better understanding of the 
backreaction of quantum energy-momentum fluctuations on the 
spacetime, the electromagnetic field might still provide suitable 
NEC violations.

The NEC violation itself is a clear consequence of quantum field 
theory and is not an assumption
(though one could reasonably question the applicability of quantum
field theory on systems with horizons.) There seems to be little doubt
that  large amplitude NEC violations do occur, though they are far 
less frequent than the small amplitude violations.
The idea that NEC violating fluctuations could have played an important
cosmological role is also to be found in the ``eternal inflation''
scenario \cite{Borde:1997pp}. While we may not be able to test the 
idea of cosmological NEC violating fluctuations, we can certainly
test quantum fluctuations with and without horizons in laboratory
experiments
\cite{Unruh:1980cg,Fischer:2004bf,Jacobson:1998he,Vachaspati:2004wn}.

The third assumption concerns the spatial extent of the fluctuation. 
Based on work done on the possibility
of creating a universe in a laboratory, topological inflation, and
wormholes, we have argued for the conjecture that small scale violations
of NEC can only give rise to universes that are affected by signals
originating at a singularity. Hence predictability is lost in such
universes. Our assumption is that even if we did know how to handle the
spacetime singularities affecting these universes, they would turn
out to be unsuitable for matter genesis. Without this assumption,
we should also be considering such universes as possible homes.

The fourth assumption is that the NEC-violating fluctuation ends in
a thermal state. All the different energy components are
also assumed to be in thermal equilibrium. We have then assumed that the
critical temperature needed for observers to exist is the temperature
at which matter-genesis occurs. One could relax this assumption
but one would need an adequate characterization of the most likely state
to be able to calculate cosmological observables (for instance, the 
spectrum of density fluctuations).

This brings us to the part of the model where we argue that even
if the large amplitude fluctuations are infrequent, they are the
only ones that are relevant for observational cosmology. This is
quite similar to the arguments given in the context of eternal
inflationary cosmology where thermalized regions are relatively
rare but these are the only habitable ones. It also occurs in chaotic
inflation \cite{Linde:1983gd}, where closed universes of all
sizes and shapes are produced but only a few are large and 
homogeneous enough to develop into the present universe. So 
this part of Island Cosmology is no weaker (and harder
to quantify) than other cosmological models.

\section{Conclusions}
\label{conclusions}

To conclude, we have described a new cosmological model, which
we call ``Island Cosmology'', where large NEC violating quantum
fluctuations (``upheavals'') in a cosmological constant filled de Sitter
spacetime create islands of matter. In Island Cosmology, spacetime
may be non-singular and eternal, and an inflationary stage is not necessary.

Like inflation, the only observational signature of Island Cosmology 
is the spectrum of density perturbations. Assuming that the spectrum 
of perturbations in Island Cosmology agrees with observations, the 
question would arise as to whether it is possible to somehow distinguish 
Island Cosmology from an inflationary scenario that is also consistent 
with observations. Unfortunately the answer seems to be  that no 
cosmological observation can distinguish between the scenarios because
of the immense adaptability of inflationary models. The
only distinguishing feature would have to come from the field
theory side since Island Cosmology does not rely on constructing
a suitable potential for a scalar field whereas this seems
to be a crucial feature in inflationary 
models. In fact, if the electromagnetic
field is subsequently determined to be capable of providing suitable NEC
violations, scalar fields might be dispensed off entirely in
Island Cosmology. The converse of this is that if the density 
fluctuations turn out not to agree with observations, we can dismiss 
the scenario of our universe being an island in the $\Lambda$-sea,
even though quantum field theory predicts the existence of such
islands. This would be an interesting conclusion as well.

Island Cosmology is attractive because it is a minimalistic model. 
It uses currently observed features of the Universe as its ingredients 
and combines them with well-established results from quantum field 
theory to account for the Universe that we live in now.

\acknowledgement
{This work was supported by the DOE and NASA at Case
Western Reserve University
}



\end{document}